\documentclass[%
 reprint,
 amsmath,amssymb,
 aps, prl,
 longbibliography, superscriptaddress 
]{revtex4-1}

\usepackage[margin=2.5cm]{geometry}
\usepackage[english]{babel}
\usepackage{graphicx}
\usepackage{dcolumn}
\usepackage{bm}
\usepackage{multirow}
\usepackage{siunitx}
\usepackage{booktabs, bigdelim, makecell}%
\usepackage{array}
\usepackage[export]{adjustbox}
\usepackage{amsmath}
\usepackage{amssymb}
\usepackage[utf8]{inputenc}

\newcommand{\im}{\mathrm{i}}
\newcommand{\mb}{\mathbf}
\usepackage{xcolor}

\usepackage{nicefrac}

\usepackage{rotating}
\usepackage[plainpages=false,pdfpagelabels,colorlinks=true,linkcolor=black,urlcolor=black,citecolor=black]{hyperref}
\usepackage{braket}

\usepackage{color, soul}

\usepackage[label font=bf,labelformat=simple, font=large, caption=false]{subfig}

\usepackage{floatrow}
\usepackage[labelformat=empty, justification=justified, format=plain]{caption}

\usepackage{caption}
\usepackage{ragged2e}
\DeclareCaptionJustification{justified}{\justifying}
\begin{document}

\title{Observing localisation in a 2D quasicrystalline optical lattice}
\author{Matteo Sbroscia}
\email{current address: Jet Propulsion Laboratory, California Institute of Technology, 4800 Oak Grove Drive, Pasadena, CA, 91109, USA}
\affiliation{Cavendish Laboratory, University of Cambridge, J.~J.~Thomson Avenue, Cambridge CB3~0HE, United Kingdom}
\author{Konrad Viebahn}
\affiliation{Cavendish Laboratory, University of Cambridge, J.~J.~Thomson Avenue, Cambridge CB3~0HE, United Kingdom}
\affiliation{Institute for Quantum Electronics, ETH Z\"{u}rich, 8093 Z\"{u}rich, Switzerland}
\author{Edward Carter}
\affiliation{Cavendish Laboratory, University of Cambridge, J.~J.~Thomson Avenue, Cambridge CB3~0HE, United Kingdom}
\author{Jr-Chiun Yu}
\affiliation{Cavendish Laboratory, University of Cambridge, J.~J.~Thomson Avenue, Cambridge CB3~0HE, United Kingdom}
\author{Alexander Gaunt}
\affiliation{DeepMind, 6 Pancras Square, London N1C 4AG, United Kingdom}
\author{Ulrich Schneider\textsuperscript{1,}}
\email{uws20@cam.ac.uk}
\date{\today}

\begin{abstract}
Quasicrystals are long-range ordered but not periodic, representing an interesting middle ground between order and disorder. We experimentally and numerically study the ground state of non- and weakly-interacting bosons in an eightfold symmetric quasicrystalline optical lattice. We find extended states for weak lattices but observe a localisation transition at a lattice depth of $V_0=1.78(2)\,E_{\mathrm{rec}}$ for the non-interacting system. We identify this transition by measuring the timescale required for adiabatic loading into the lattice, which diverges at the critical lattice depth for localisation. 
Gross-Pitaevskii simulations show that in interacting systems the transition is shifted to deeper lattices, as expected from superfluid order counteracting localisation. Our experimental results are consistent with such a mean-field shift. Quasiperiodic potentials, lacking conventional rare regions, provide the ideal testing ground to realise many-body localisation in 2D.
\end{abstract}

\maketitle
Quasiperiodic potentials give rise to fractal, self-similar structures both in momentum space~\cite{senechal_quasicrystals_1995} and in their energy spectrum~\cite{asch_cantor_2006}. The unique and intriguing status of quasiperiodicity as a middle ground between order and disorder has generated interests in a wide range of fields, from recent experiments with twisted bilayer graphene \cite{ahn_dirac_2018}, topological photonics \cite{vardeny_optics_2013}, and ultracold atoms~\cite{guidoni_quasiperiodic_1997, roati_anderson_2008, gadway_glassy_2011, schreiber_observation_2015, rajagopal_phasonic_2019}, to a proof for the undecidability of the spectral gap theorem \cite{cubitt_undecidability_2015}.
Combining two-dimensional (2D) quasicrystalline potentials with ultracold atoms will not only enable the observation of the 2D Bose glass~\cite{fisher_boson_1989, soyler_phase_2011} but could prove essential for deciding the ultimate fate of many-body localisation (MBL) in 2D. 
Though established and confirmed in 1D \cite{schreiber_observation_2015, Andp2017}, the existence of MBL in higher dimensions is --- despite first experimental studies~\cite{Choi2016,Bordia2D} --- less clear, as rare ergodic enclosures might trigger a thermalisation avalanche that destabilises the localised state \cite{de_roeck_stability_2017,Potirniche2019}. In quasiperiodic systems, in contrast, the long-range ordered nature of the potential precludes such conventional rare ergodic regions~\cite{Szabo2020}.
\vspace*{2.7mm}

By studying the diverging timescale required for adiabatically loading a Bose-Einstein condensate (BEC) of  $\sim 3 \times 10^5$ $^{39}$K atoms into a quasicrystalline optical lattice, we probe the disorder-induced localised phase and its resilience against interactions. We perform the investigation in momentum space by recording matterwave diffraction patterns, instead of employing the real space in-situ techniques mostly used so far.
We create a quasicrystalline 2D optical lattice $V(\mathbf{r})$ by superimposing four coplanar 1D lattices formed by retro-reflected laser beams (wavelength $\lambda_{\mathrm{lat}}=725\,$nm) under $45^\circ$ angles and mutually detuned by at least $\SI{10}{MHz}$, as illustrated in Fig.\ \ref{fig:IPR}a:
\begin{equation}
V(\mathbf{r}) = V_0 \textstyle \sum_{i=1}^4 \sin^2(\mathbf{k}_i\cdot\mathbf{r}),
\end{equation} 
\begin{equation}\label{eq:defs}
\mathbf{k}_i\in\frac{2\pi}{\lambda_{\mathrm{lat}}}\cdot 
\left\{\begin{pmatrix} 1\\0\end{pmatrix},\:
\frac{1}{\sqrt{2}}\begin{pmatrix} \pm1\\1\end{pmatrix},\:
\begin{pmatrix} 0\\1\end{pmatrix}\right\}.  
\end{equation}
Analogously to solid-state quasicrystals \cite{shechtman_metallic_1984} this potential exhibits a crystallographically forbidden eightfold rotational symmetry~\cite{viebahn_matter-wave_2019}. Fig.\ \ref{fig:IPR}b  shows the numerically calculated single-particle ground state of this potential for different lattice depths $V_0$. In contrast to periodic lattices, there exists a critical lattice strength $V_{\mathrm{loc}}$ at which the ground state undergoes a transition from an extended, quasiperiodic Bloch wave to an exponentially localised state. The quasiperiodic potential generically possesses a \emph{single} absolute minimum and the absolute ground state in the localised phase is centred around this site. 

Since disorder-induced localised phases cannot be detected in equilibrium, we rely on a dynamical probe: starting with a non-interacting BEC in a dipole trap (see~\cite{viebahn_matter-wave_2019} for details), we increase the lattice depth linearly from $0$ to $V_0$ in a time $\tau$ before decreasing it back to $0$ using the same ramp and performing time-of-flight imaging, as illustrated in Fig.\ \ref{fig:IPR}c. Whenever the loading and unloading dynamics under this triangular ramp are fully adiabatic, all atoms return to the initial BEC wavefunction and we detect them all at zero momentum.

\begin{figure*}[t!]
\centering
\includegraphics[scale=1]{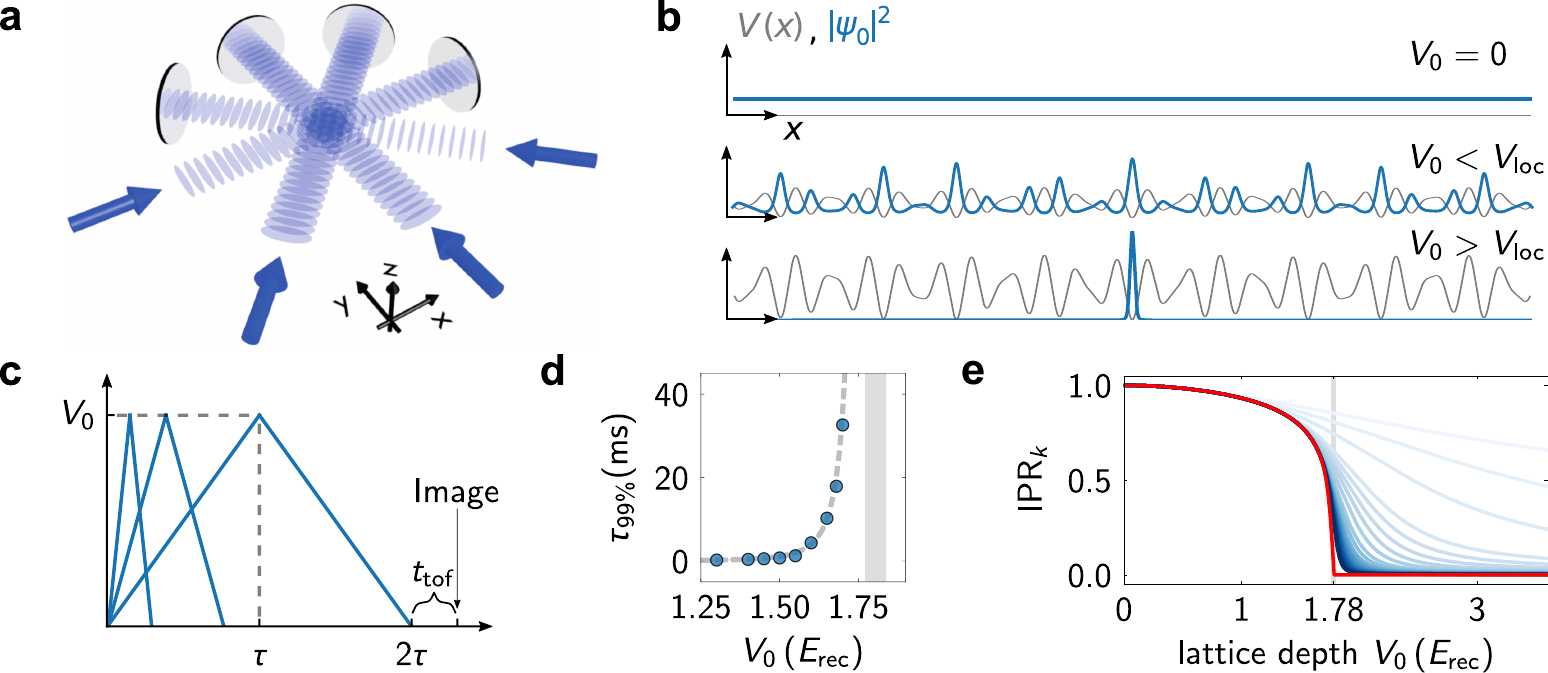}
\caption{\textbf{Fig.\ 1\,\textbar\,Localisation in a 2D quasicrystalline optical lattice.} \textbf{a}, Eightfold symmetric quasicrystalline optical lattice formed by superimposing four standing waves. \textbf{b}, Numerically computed single-particle ground states for lattice depths below $(V_0 / E_{\mathrm{rec}} = 0, 1)$ and above $(V_0 / E_{\mathrm{rec}}=2)$ the localisation transition. Grey lines denote the lattice potential. \textbf{c}, The lattice ramp used in all experiments. The final momentum distribution is detected using a time-of-flight of $t_{\mathrm{tof}}=33\,$ms. \textbf{d}, Numerically calculated ramp time (circles) needed for a $99\%$ recovery of the zero-momentum peak population in the quasicrystalline lattice. This timescale diverges at a lattice depth (grey bar) compatible with the localisation transition ($V_{\mathrm{loc}}$). The dashed line denotes a phenomenological fit \cite{noauthor_notitle_nodate-1}. \textbf{e}, Calculated Inverse Participation Ratio of the single-particle ground state in momentum space (IPR$_k$), exhibiting a cusp at the critical lattice depth $V_{\mathrm{loc}} \approx 1.78(2)\,E_{\mathrm{rec}}$ (grey bar). 
Darker blue lines correspond to larger basis sets; the red line marks the extrapolation to the infinite basis, see supplemental material \cite{noauthor_notitle_nodate-1}.}
\label{fig:IPR}
\end{figure*}
\begin{figure}[b!]
\vspace*{-0.15cm}
\setcounter{figure}{1}
\centering
\includegraphics[scale = 1]{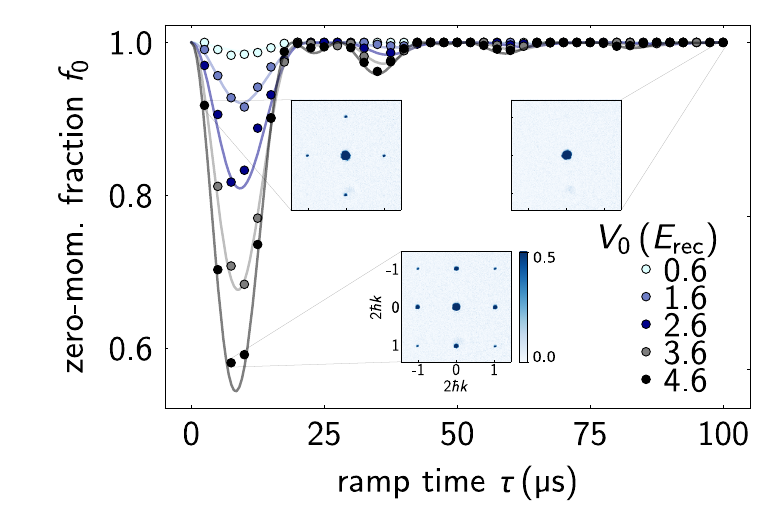}%
\vspace*{-0.25cm}
\caption{\textbf{Fig.\ 2\,\textbar\,Adiabatic triangular lattice ramps for a regular square lattice.} The zero-momentum fraction $f_0$ is fully recovered for all probed lattice depths already for rather short ramp durations of $\tau \gtrsim 40\,\si\micro\mathrm{s}$. Different colors correspond to different maximum lattice depths $V_0$; solid lines denote numerical solutions to the single-particle Schr\"{o}dinger equation~\cite{noauthor_notitle_nodate-1}. Insets show time-of-flight images for $V_0=4.6\,E_{\mathrm{rec}}$ and different ramp times $\tau$: the optical density colorscale is chosen to highlight small excited populations, thereby saturating the zero-momentum peak.}
\label{fig:2D}
\end{figure}
\begin{figure*}[t!]
\setcounter{figure}{2}
\centering
\includegraphics[scale=1]{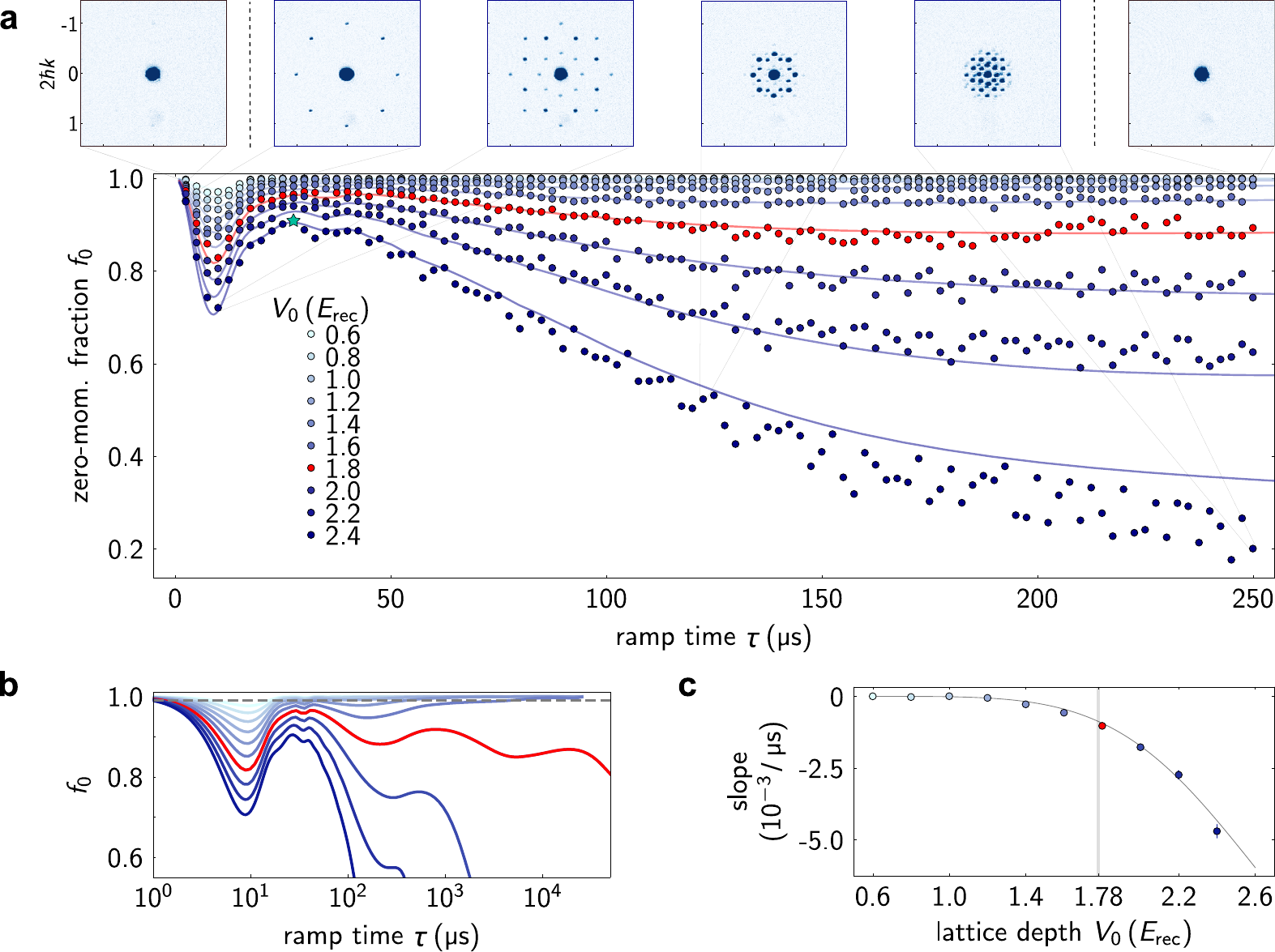}
\caption{\textbf{Fig.\ 3\,\textbar\,Breakdown of adiabaticity in a quasicrystalline lattice.} \textbf{a}, Final population of the zero-momentum state for the quasicrystalline lattice. Solid lines denote the numerical solution to the single-particle Schr\"odinger equation~\cite{noauthor_notitle_nodate-1}, also shown in \textbf{b} for longer ramp times. Both experiment and simulation show that the zero-momentum fraction $f_0$ ultimately recovers to one for $V_0 < V_{\mathrm{loc}}$, but tends to zero for $V_0 > V_{\mathrm{loc}}$. The red data denotes the lowest probed lattice depth above the localisation transition, which is also the first one not to recover. The turquoise star denotes the maximum recovery for $2.4\,E_{\mathrm{rec}}$. The first and last insets in \textbf{a} show time-of-flight images for $0.6\,E_{\mathrm{rec}}$, all others are for $2.4\,E_{\mathrm{rec}}$; the colorscale is identical to Fig.\ \ref{fig:2D}.  The dashed line in \textbf{b} denotes $99\%$ recovery, the ramp time required to reach this is plotted in Fig.\ \ref{fig:IPR}d. \textbf{c}, Fitted slopes for $\tau = 50-\SI{100}{\micro s}$: large negative slopes indicate localisation.
}
\label{fig:4D}
\end{figure*}

For $V_0<V_{\mathrm{loc}}$, the lattice ground state is a quasiperiodic but extended wave (Fig.\ \ref{fig:IPR}b) and the initially constant density only has to rearrange locally. Similarly to a periodic lattice~\cite{denschlag_bose-einstein_2002} (discussed later), the BEC can hence be adiabatically loaded in a finite time. Above the localisation transition at $V_{\mathrm{loc}}$, on the other hand, the initially extended ground state wavefunction  becomes exponentially localised around the typically unique global minimum and is characterised by a rapidly shrinking localisation length. Following this change adiabatically would require particle transport over large distances, leading to a diverging adiabaticity timescale in the thermodynamic limit. This prediction is confirmed by the numerical simulation shown in Fig.\ \ref{fig:IPR}d: the ramp time $\tau$ needed to return $>99\%$ of the atoms to the zero-momentum state diverges upon approaching the localisation transition~\cite{noauthor_notitle_nodate-1}. In order to precisely pinpoint the critical lattice depth, we numerically calculate the single-particle ground state using a plane wave basis and compute its Inverse Participation Ratio in momentum space $\mathrm{IPR}_k = \sum_j |\tilde{\psi_j}|^4$, where $\tilde{\psi}_j$ denotes the amplitude of the $j$\textsuperscript{th} plane wave \cite{noauthor_notitle_nodate-1}. As shown in Fig.\ \ref{fig:IPR}e, this displays a sharp transition at $V_{\mathrm{loc}} = 1.78(2)\,E_{\mathrm{rec}}$, which is consistent with a numerical calculation of the localisation transition in real space~\cite{Szabo2018}. 

In our experiments, we first focus on non-interacting atoms (scattering length $a \approx 0\,a_0$ \cite{fletcher_two-_2017}) and a regular square lattice generated by two of the four standing waves. In this periodic case there is no localisation transition and adiabatic loading should thus always be possible~\cite{denschlag_bose-einstein_2002}. 
We individually fit all momentum peaks in the time-of-flight images taken at the end of the lattice ramp and extract both the total number of atoms $N$ and the atom number in the zero-momentum peak $N_0$. Hence we define $f_0 = N_0/N$ as the zero-momentum fraction, plotted in Fig.\ \ref{fig:2D} as a function of the ramp time $\tau$. For $\tau=0$ the atoms have no time to respond and they simply remain in the zero-momentum state, hence $f_0=1$. For small but finite ramp times, which mean rather abrupt lattice changes, the non-adiabatic dynamics gives rise to significant excitations that are strongest for ramp times around $\tau\approx 10\,\si\micro$s. For slower ramps, the atoms begin to adiabatically follow the ground state and return to the initial state, eventually leading to a full recovery of the zero-momentum peak population for all inspected lattice depths. The oscillatory behaviour at intermediate ramp times stems from only a few discrete momenta being relevant to the dynamics~\cite{denschlag_bose-einstein_2002,viebahn_matter-wave_2019}. Our observations are in perfect agreement with a numerical solution of the single-particle Schr\"odinger equation for a homogeneous lattice without any trap (solid lines in  Fig.\ \ref{fig:2D}).
The observed timescales for adiabatic loading  ($<100\,\si\micro\mathrm{s}$) are much shorter than those typically used ($50-200\,$ms \cite{gericke_adiabatic_2007}), as the system is non-interacting and we can neglect the harmonic trap on these short timescales \cite{noauthor_notitle_nodate-1}.

In the quasicrystalline lattice, in contrast, the dynamics at long ramp times crucially depends on whether the maximum lattice depth lies below or above the localisation transition at $V_{\mathrm{loc}} = 1.78(2)\,E_{\mathrm{rec}}$. As shown in Fig.\ \ref{fig:4D}a, the dynamics starts similarly to the periodic case, also giving rise to a maximum depletion around $\tau\approx 10\,\si\micro\mathrm{s}$. 
For longer ramps, however, the quasiperiodic character leads to two dramatically distinct behaviours: while for low lattice depths the zero-momentum peak eventually fully recovers (see also Fig.\ \ref{fig:4D}b), it never does for $V_0 > V_{\mathrm{loc}}$. Its population instead keeps decreasing and ultimately tends to zero for slower ramps, where atoms can reach higher diffraction orders. These include states at successively lower momenta and lead to the atoms ``getting lost'' in the fractal momentum space of the quasicrystal~\cite{viebahn_matter-wave_2019}. This is directly visible in the time-of-flight pictures at the top of Fig.\ \ref{fig:4D}a, where the populated momenta are situated closer to the origin for longer $\tau$. A numerical solution of the single-particle Schr\"odinger equation (solid lines in Fig.\ \ref{fig:4D}a) not only agrees very well with the experimental data, but furthermore allows exploration of much slower ramps without trap effects (Fig.\ \ref{fig:4D}b), confirming the two distinct dynamics for slow ramps. All traces show log-periodic oscillations, which are a sign of discrete scale invariance and of the fractal nature of the system \cite{sornette_discrete-scale_1998}.  Fig.\ \ref{fig:4D}c shows the gradients extracted from linear fits between $\tau = 50-\SI{100}{\micro s}$. While the log-periodic oscillations can give rise to shallow negative slopes even below the transition, localisation causes stronger negative slopes that highlight the counterintuitive nature of the dynamics in this regime -- the slower the ramps, the further away from adiabaticity. The small discrepancies at long ramp times are probably caused by slightly inaccurate lattice depth calibrations.

In order to investigate the effect of repulsive interactions on the localisation transition, we applied different magnetic fields during the lattice ramp to sample scattering lengths between $0-200\,a_0$, as shown in Fig.\ \ref{fig:interactions}. 
In this rather deep lattice of $V_0 = 2.4\, E_{\mathrm{rec}}> V_{\mathrm{loc}}$ the non-interacting ground state is strongly localised, leading to a very low recovery of the zero-momentum fraction of $f_0 \approx 0.2$ for a ramp time $\tau=250\,\si\micro$s. Interestingly, adding sufficiently strong repulsive interactions increases the recovered fraction to almost unity. 
The dashed line at $f_0\approx 0.9$ denotes the recovered fraction at its maximum value for $\tau \approx \SI{30}{\micro s}$ (turquoise star in Fig.\ \ref{fig:4D}a), which is essentially independent of the interaction strength (see supplemental material~\cite{noauthor_notitle_nodate-1}). For weak interactions, the low recovered fraction after $250\,\si\micro$s therefore directly corresponds to strong negative slopes that signal localisation, whereas the close to perfect recovery for larger interactions suggest that no phase transition has been crossed. This is consistent with the expectation that repulsive interactions in a BEC shift the onset of localisation to deeper lattices. The same effect can also be observed in the Gross-Pitaevskii equation (GPE) simulations \cite{noauthor_notitle_nodate-1} shown in the main inset of Fig.\ \ref{fig:interactions}, where the superfluid fraction $\psi_{\mathrm{SF}}$ vanishes only at deeper lattices in the interacting case. 

By recording matterwave diffraction patterns after a triangular lattice ramp, we could observe the localisation transition in a 2D quasiperiodic potential, thereby establishing a novel and complementary observable to study disorder-induced localisation. The timescale required for adiabatically loading a non-interacting condensate into the lattice diverges at the localisation transition at $V_{\mathrm{loc}} \approx 1.78(2)\, E_{\mathrm{rec}}$, as the change from an extended to a localised wavefunction would require mass transport across the whole system. The localised phase is stable against adding small repulsive interactions and the observed behaviour is consistent with a mean-field shift of the transition point.
Adding a strong 1D lattice along the third direction will extend these experiments into the 2D regime and allow for strong correlations. This will provide access to the full disorder versus interaction phase diagram, including the so far elusive 2D Bose glass and
 a re-entrant transition where adding disorder turns a Mott insulator back into a superfluid \cite{soyler_phase_2011}. Localised states in 2D quasiperiodic potentials could prove crucial for studying MBL in 2D, as their long-range ordered nature ensures that conventional rare ergodic regions are absent \cite{Szabo2020}. They are also an ideal starting point to study interacting Floquet systems, where the non-ergodicity could effectively mitigate the typical Floquet heating \cite{Eckardt2017, bordia_periodically_2017}.


\begin{figure}
\setcounter{figure}{3}
\centering
\includegraphics[scale = 1]{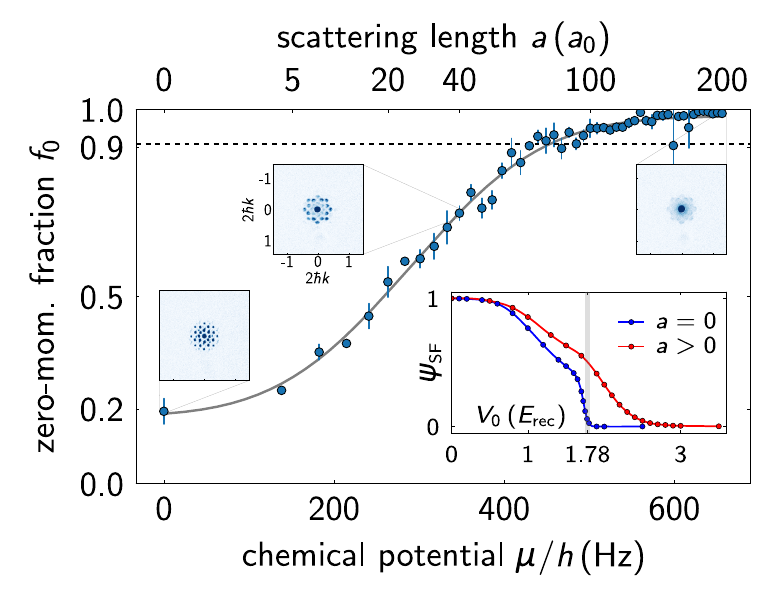}%
\vspace*{-0.25cm}
\caption{\textbf{Fig.\ 4\,\textbar\,Effect of repulsive interactions.} The zero-momentum fraction for a fixed slow ramp $\tau = \SI{250}{\micro s}$ and deep lattice $V_0 = 2.4\, E_{\mathrm{rec}}$ increases with growing interaction strengths, given by the scattering length ($a$) and the resulting chemical potential ($\mu\propto a^{2/5}$). The black solid line denotes a phenomenological fit using the error function of equation (\ref{eq:errfunc}) in \cite{noauthor_notitle_nodate-1}. Error bars denote the standard deviation of three measurements. The main inset is a mean-field GPE simulation of the superfluid stiffness or fraction $\psi_{\mathrm{SF}}$ \cite{noauthor_notitle_nodate-1} showing that the localisation transition is pushed to deeper lattices in interacting systems. The colorscale used in the small insets is identical to Fig.\ \ref{fig:2D}. The chemical potential scale has a systematic uncertainty on the order of $10-15\%$ stemming from the atom number calibration.}
\label{fig:interactions}
\end{figure}
\vfill

\newpage

\section{Acknowledgements}

We would like to thank Oliver Brix, Michael H{\"o}se, Max Melchner, and Hendrik von Raven for assistance during the construction of the experiment, and Shaurya Bhave for assistance in the final stages of the measurements. We are grateful to Zoran Hadzibabic and his team, Juan Garrahan, and Attila Szab\'o for helpful discussions. This work was partly funded by the European Commission ERC Starting Grant \mbox{QUASICRYSTAL}, the EPSRC Grant (EP/R044627/1), and Programme Grant \mbox{DesOEQ} (EP/P009565/1).  

%




\clearpage
\section{Supplemental material}

\subsection{Experimental details}

The experimental apparatus and sequence is described in the supplemental material of \cite{viebahn_matter-wave_2019}. We load $2.5-3 \times 10^5$ $^{39}$K BEC atoms from a $(\omega_x, \omega_y, \omega_z) = 2\pi\cdot(19, 18, 115)\,$Hz crossed-beam dipole trap into a $\lambda_{\mathrm{lat}} = \SI{725}{nm}$ blue-detuned lattice in the $x,y$-plane. In order to avoid interferences between the individual $45^\circ$-spaced 1D lattices, they are detuned from each other by at least $\SI{10}{MHz}$, much higher than any frequency the atoms could respond to. There is no lattice along the third direction.

All images are taken after a $\SI{33}{ms}$ time-of-flight, and $\SI{2}{ms}$ into it interactions are re-introduced to a value of $a=280\,a_0$ in order to ``puff up'' the diffraction orders and to minimise saturation of the images, which would otherwise prevent a reliable fitting procedure. The lack of visible scattering spheres in the images demonstrates that the effects of the ``puffing up'' on the dynamics is negligible.

We can neglect the effects of the harmonic trap, as they would start to dominate only after times on the order of $1/\omega_{x,y}\approx 10\,$ms. 

\subsection{Population extraction}

The populations of the individual momenta are extracted by fitting a Thomas-Fermi profile to each peak independently. 
A fit is performed instead of a pixel count as it allows for a reliable characterisation of the peaks even in the presence of noise and (more importantly) saturation. Despite the ``puffing up'' procedure, the zero-momentum peak (the original BEC) remains almost always severely saturated. Only points below a  cutoff of OD$=3.4$ are considered in the fit, OD being the optical density.

For higher diffraction orders, the fits are generally reliable for the square and for the short-term dynamics of the eightfold quasiperiodic lattices, where the peaks are well separated. For the longest and deepest quasiperiodic lattice ramps, on the other hand, atoms start to occupy up to the $10$\textsuperscript{th} diffraction order, such that peaks begin to overlap in the images. 


The color scale chosen for all insets presented in the paper was capped at OD $=0.5$ in order to emphasise small excited populations. This choice in turn visually over-emphasises the spatial extent and hence the population of the zero-momentum peak, which in the deeply localised regions is highly depleted. As shown in Extended Data Fig.\ \ref{fig:4Dintimages}, different colorscales visually confirm that higher interaction strengths result in an almost complete recovery of the zero-momentum peak.

\begin{figure}
\setcounter{figure}{0}
\includegraphics[scale = 1]{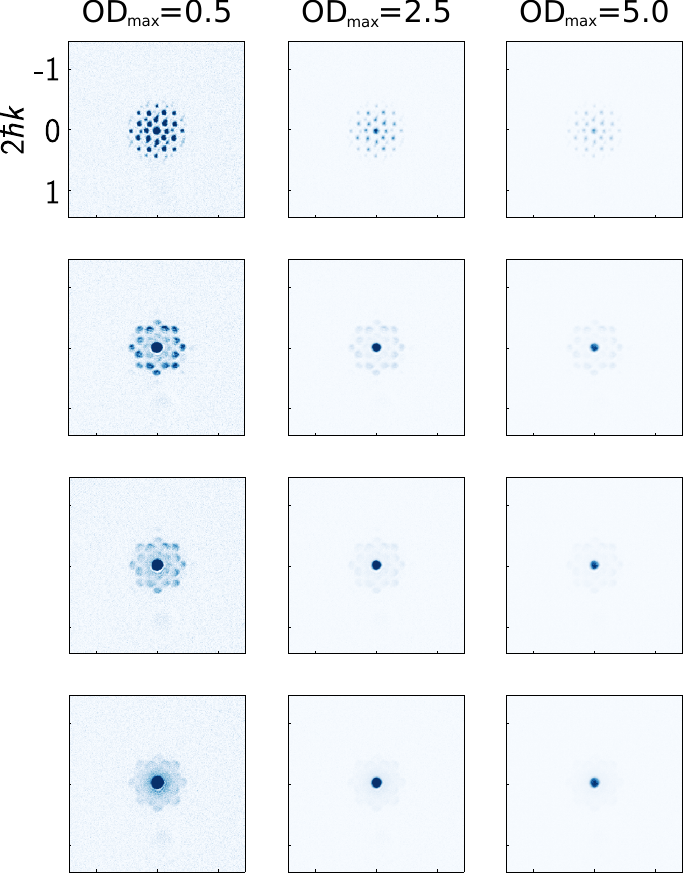}%
\caption{\textbf{Extended Data Fig.\ 1\,\textbar\,Different colorscales.} The four different rows, starting from the top one, correspond to $a/a_0 = 0, 40, 100,$ and $200$. }
\label{fig:4Dintimages}
\end{figure}

\subsection{Simulations}


\subsubsection{Momentum space}

The numerical data presented in Fig.\ \ref{fig:IPR}e and Fig.\ \ref{fig:2D} was produced by solving the single-particle problem using a momentum space representation. 
As the four 1D lattices (depicted in Fig.\ \ref{fig:IPR}a) are detuned from each other by at least $\SI{10}{MHz}$, there is no relevant cross interference and the main interaction consists of atoms absorbing a photon from one laser beam and remitting it into the counter-propagating beam. This changes the atomic momentum by $2\hbar \mathbf{k}$. The wavevectors of the four incoming beams have the same magnitude $|\mathbf{k}|=2\pi/\lambda_{\mathrm{lat}}$ and only differ in their direction, as listed in equation\ (\ref{eq:defs}).

Employing the language of synthetic dimensions \cite{Ozawa2019}, the four 2D wavevectors can be chosen as orthogonal vectors in a four dimensional space: 
\begin{align}
\mathbf{k}_1 = \begin{pmatrix} 1\\0\end{pmatrix} \rightarrow \begin{pmatrix} 1\\0\\0\\0\end{pmatrix}, \quad \mathbf{k}_2 = \frac{1}{\sqrt{2}}\begin{pmatrix} 1\\1\end{pmatrix} \rightarrow \begin{pmatrix} 0\\1\\0\\0\end{pmatrix}, \\
\mathbf{k}_3 = \frac{1}{\sqrt{2}}\begin{pmatrix} -1\\1\end{pmatrix} \rightarrow \begin{pmatrix} 0\\0\\1\\0\end{pmatrix}, \quad \mathbf{k}_4 = \begin{pmatrix} 0\\1\end{pmatrix} \rightarrow \begin{pmatrix} 0\\0\\0\\1\end{pmatrix},
\end{align}
where the common factor $2\pi/\lambda_{\mathrm{lat}}$ has been omitted.

Starting from a zero-momentum state ($\mathbf{q}=0$), the atoms can then reach the following momenta $\mb{b}_p$  (matterwave diffraction vectors) by two-photon transitions:
\begin{equation}\label{eq:basis}
\mb{b}_p/2 = i\mathbf{k}_1 + j\mathbf{k}_2 + \ell\mathbf{k}_3 + m\mathbf{k}_4, \quad (i, j, \ell, m) \in \mathbb{Z}^4.
\end{equation}
Each coefficient represents the number of two-photon transitions along one of the standing waves (1D lattices). The blue lines in Fig.\ \ref{fig:IPR}e correspond to the following truncations:
\begin{equation}\label{eq:n}
|i|+|j|+|\ell|+|m|\leqslant n\quad\textrm{for}\quad n=1-25. 
\end{equation}
The single-particle $d$-dimensional Hamiltonian for $\mathbf{q}=0$, where $d$ is the number of active lattice beams and hence the dimensionality of the augmented space, is then given by:
\begin{eqnarray} \label{eq:hamiltonian}
H^d_{p,q}(t) = \begin{cases}
E_{\mathrm{rec}} \times |\mathcal{P}^\parallel_d \cdot \mb{b}_p|^2+ d \frac{V_0(t)}{2}\quad \text{for} \quad p = q \quad\\
 V_0(t)/4\quad \text{for} \quad || p-q || = 1\\
0 \quad \text{otherwise,}
 \end{cases},
\end{eqnarray}
where $\mathbf{b}_p$ is the basis vector of equation (\ref{eq:basis}) with $p=(i,j,\ell,k)$ and $\mathcal{P}^\parallel$ is the projection matrix from the augmented back to the physical 2D basis. For the regular square lattice $d=2$, while for the eightfold quasicrystalline lattice $d=4$, and the projection matrix $\mathcal{P}^\parallel_4$ is 
\begin{equation} \label{eq:projmatrix4D}
\mathcal{P}^\parallel_4 = \left (  \begin{array}{cccc}
1 & \cos\nicefrac{\pi}{4} & -\cos\nicefrac{\pi}{4} & 0 \\
0 & \sin\nicefrac{\pi}{4} & \sin\nicefrac{\pi}{4} & 1 \\
0 & 0 & 0 & 0 \\
0 & 0 & 0 & 0
\end{array} \right ).
\end{equation}

We determine the ground state of $H$ using Lanczos' algorithm and calculate the ground state IPR$_k$ shown in Fig.\ \ref{fig:IPR}e.

The dynamical simulations of Fig.\ \ref{fig:2D} were obtained by direct integration of the dimensionless Schr{\"o}dinger equation:
\begin{equation}
\im \frac{\mathrm{d}}{\mathrm{d}\tau_{\mathrm{rec}}}|\psi\rangle = H'|\psi\rangle,
\end{equation}
where $H' = H/E_{\mathrm{rec}}$ and $\tau_{\mathrm{rec}} = \hbar/E_{\mathrm{rec}}$.

The simulation requires a truncation in the plane wave basis, which limits the reliability of its predicted dynamics near the localisation transition because of the large number of Fourier components involved. Hence, a real space simulation was employed in these regions (next section).

\subsubsection{Real space}

The numerical data in Fig.\ \ref{fig:IPR}b, Fig.\ \ref{fig:IPR}d, Fig.\ \ref{fig:4D}, and in the inset of Fig.\ \ref{fig:interactions} was produced 
using the dimensionless Gross-Pitaevskii equation (GPE) in real space. It was solved using the split-step Fourier method \cite{gaunt_degenerate_2015} in the presence of an external potential energy term:
\begin{equation} \label{eq:GPEalex}
\im\frac{\partial \psi'}{\partial \tau_{\mathrm{rec}}} = -\nabla_\xi^2 \psi' + \alpha V(\boldsymbol{\xi})\psi' + 8 \pi a' |\psi'|^2\psi',
\end{equation}
with $\boldsymbol{\xi} = k_{\mathrm{lat}}\mb{x}$, $a' = a k_{\mathrm{lat}}$ and $\psi' = \psi / k_{\mathrm{lat}}^{3/2}$. The lattice depth in units of $E_{\mathrm{rec}}$ is $\alpha$. The non-interacting (single-particle) case was realised by setting $a'=0$. 
The use of this mean-field calculation is further justified by the absence of an additional lattice along the vertical direction, resulting in a weak confinement along that direction and therefore weak correlations. 

\begin{figure}
\centering
\includegraphics[scale = 1]{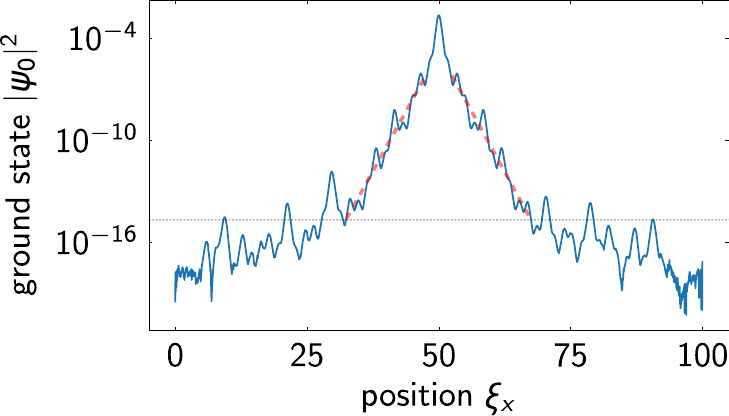}%
\caption{\textbf{Extended Data Fig.\ 2\,\textbar\,Exponentially decaying wavefunction in the localised state.} The GPE of equation (\ref{eq:GPEalex}) was evolved in imaginary time for $\alpha=2$ to find the (localised) ground state, whose density profile $|\psi_0|^2$ is plotted here as a function of the dimensionless coordinate $\xi_x$. The red dashed lines show the characteristic exponential localisation around the global minimum (chosen to lie at $\xi_x=50$ here). The dashed grey line denotes where the simulation stops being reliable due to float precision and finite evolution time, signalled by the loss of even symmetry.}
\label{fig:loc}
\end{figure}

This equation was evolved in imaginary time to find the (localised) ground state for $\alpha = 2$, shown in Fig.\ \ref{fig:IPR}b and in Extended Data Fig.\ \ref{fig:loc} in log-scale. For the dynamical simulations of Figs.\ \ref{fig:4D}a and \ref{fig:4D}b, it was instead evolved in real time with a given ramp $V_0(t)$ from a uniform initial wavefunction. 

The ramp times $\tau_{99\%}$ in Fig.\ \ref{fig:IPR}d were fitted with:
\begin{equation}\label{eq:fitpar}
y = \frac{a}{\left ( V_{\mathrm{loc}} - V_0 \right)^b}, \quad \mathrm{with} \quad 
\begin{tabular}{c|c|c|}
 & value & error\\
\hline
$a$ & 0.05 & 0.04 \\
\hline
$V_{\mathrm{loc}} $& 1.80 & 0.03\\
\hline
$b$ & 3.15 & 0.80\\ 
\hline
\end{tabular}
\end{equation}
which provided an estimate for $V_{\mathrm{loc}}/E_{\mathrm{rec}} = 1.80(3)$, shown as the grey bar. 

For the inset in Fig.\ \ref{fig:interactions}, the simulation solved for the ground state of the lattice of equation (\ref{eq:GPEalex}) in the presence of twisted boundary conditions~\cite{lieb_superfluidity_2002}, where the phase of the wavefunction is required to differ by a small angle $\theta$ between opposite boundaries of the simulation region. A wavefunction localised at the origin is unaffected by such twist, as its magnitude at the boundary vanishes: the ground state energy of the localised phase is hence independent of the twist. A delocalised wavefunction, in contrast, displays a phase difference $\theta$ over the system size $L$, corresponding to a phase velocity $v \propto \theta/L$ and an associated additional kinetic energy.
The sensitivity of the ground state energy between the untwisted and twisted systems ($\Delta E_{\mathrm{g}}/\theta^2$) is directly proportional to the superfluid stiffness or superfluid fraction $\psi_{\mathrm{SF}}$  and is plotted in the inset of Fig.\ \ref{fig:interactions} as a function of lattice depth. At the localisation transition, the superfluid fraction vanishes. 

To quantify the strength of the interactions, we consider the spatially averaged interaction energy per atom in a 3D box of size $L$:
\begin{equation}
E_{\mathrm{int}}  = \frac{1}{L^3} \int \frac{4 \pi \hbar^2 a}{m}  n(\mb{r}) \,\mathrm{d}^3\mathbf{r},
\end{equation}
where $n(\mb{r})$ is the 3D density of the system. Assuming a system with atom number $N$ that is translationally invariant in $z$, the simulated wavefunction $\psi$ is normalised such that $\int |\psi|^2\, \mathrm{d}^3\mathbf{r} = L \int  |\psi|^2 \,\mathrm{d}x\, \mathrm{d}y = \eta L,$ where we defined $\eta = N / L$. This leads to:
\begin{equation}
E_{\mathrm{int}}  = \frac{1}{L^2}\frac{4 \pi \hbar^2 a}{m} \eta,
\end{equation}
and substituting in $E_{\mathrm{rec}} = \hbar^2 k^2 / (2m)$ gives 
\begin{equation}
E_{\mathrm{int}} = 8  \pi  a \frac{ \eta }{k^2 L^2} E_{\mathrm{rec}}.
\end{equation}
For the red curve in the inset of Fig.\ \ref{fig:interactions}, a value of $a  \eta = 0.3$ was used in a simulation volume of size $k L = 14 \pi$, which corresponds to $E_{\mathrm{int}} = 0.004 E_{\mathrm{rec}}$.
We checked that performing the simulation in a volume of size $k L = 64 \pi$ while preserving $E_{\mathrm{int}}$ does not qualitatively affect the results.

\subsubsection{Inverse Participation Ratio in momentum space (IPR$_k$)}

A quantum state may be expressed in any complete orthonormal basis, for instance in the Wannier $|w_i\rangle$ or in the plane wave $|\phi_j\rangle$ bases:
\begin{equation}
|\Psi\rangle = \sum_i \psi_i |w_i \rangle = \sum_j \tilde{\psi}_j |\phi_j\rangle.
\end{equation}  
The Inverse Participation Ratio in momentum space $\mathrm{IPR}_k$ is defined as
\begin{equation}\label{eq:iprpr}
\mathrm{IPR}_k = \sum_j |\tilde{\psi_j}|^4, 
\end{equation}
and measures how many plane waves contribute to a given state. A large and finite $\mathrm{IPR}_k$ corresponds to the situation where only a handful of plane waves are of relevance --- a large IPR is equivalent to a small participation ratio (PR), meaning a small ratio of participating states to available states. 
Normalisation imposes $\sum_i |\tilde{\psi}_i|^2 = 1$, and hence $\sum_i |\tilde{\psi}_i|^4 \leqslant 1$. The equality is only satisfied when the state consists of only \emph{one} basis function, i.e.\ for a single plane wave. A state that is localised in real space, on the other hand, necessarily comprises an infinite number of planes waves and hence will have a vanishing $\mathrm{IPR}_k$.

For periodic lattices, the single-particle eigenstates are always extended Bloch waves with a finite $\mathrm{IPR}_k$, which approaches zero only asymptotically with lattice depth. The eightfold quasicrystalline lattice, in contrast, displays the sharp transition shown in Fig.\ \ref{fig:IPR}e, signalling the localisation transition.
 

The following table provides a summary for the ground state IPR in real ($\mathrm{IPR}_x$) and momentum ($\mathrm{IPR}_k$) space characterising the delocalised (extended) and localised phases, highlighting the opposite behaviours:

\begin{table}[H]
\begin{tabular}{|c|c|c|c|}
\hline
 & & \makecell{delocalised in $x$ \\ (localised in $k$)} & \makecell{localised in $x$ \\ (delocalised in $k$)}   \\ 
\hline
$\mathrm{IPR}_x$ & $\sum_i |\psi_i|^4$ & \makecell{$\rightarrow 0$ \\ (thermo. limit)} &  \makecell{finite \\ ($1$, Wannier functions)} \\
\hline
$\mathrm{IPR}_k$ &  $\sum_j |\tilde{\psi_j}|^4$ & \makecell{finite \\ ($1$, no lattice)}  & \makecell{$\rightarrow 0$ \\ (thermo. limit)}  \\
\hline
\end{tabular}
\end{table}

The IPR$_k$ of the first few tens of low-lying eigenstates was also computed and exhibited the same qualitative behaviour as the ground state. The red curve in Fig.\ \ref{fig:IPR}e was obtained by extrapolating the blue curves to the $n\rightarrow \infty$ limit --- see equation (\ref{eq:n}). To this end we plotted the computed data for each lattice depth against $1/n$ and extracted the intercept of a fitted straight line.

\subsection{Interacting data}

The interaction strength was varied by employing the Feshbach resonance at $402.70(3)$ G of the $|1,1\rangle$ state in $^{39}$K \cite{fletcher_two-_2017}.

We note that the maximum recovery of $f_0\approx 0.9$ at $\tau \approx \SI{30}{\micro s}$ (turquoise star in Fig.\ \ref{fig:4D}a) for $V_0 = 2.4\,E_{\mathrm{rec}}$ is independent of the interaction strength in the investigated range: the chemical potential at $a=200\,a_0$ in our harmonic trap is $\mu/h \approx 742 \pm \SI{84}{Hz}$. This corresponds to an interaction timescale $\sim \SI{1.5}{ms}$, which is large compared to the $\tau \approx \SI{30}{\micro s}$ recovery time.

\begin{figure}
\vspace*{0.15cm}
\centering
\includegraphics[scale = 1]{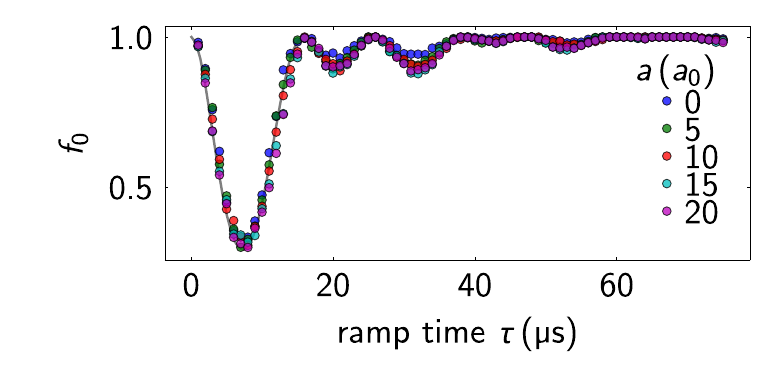}%
\caption{\textbf{Extended Data Fig.\ 3\,\textbar\,Absence of interaction effects on square lattice ramps.} The zero-momentum fraction $f_0$ is fully recovered for square lattices even for moderate repulsive interaction strengths, quantified by the scattering length $a$. Here, $V_0 = 7\,E_{\mathrm{rec}}$. The solid line is the same numerical solution to the single-particle Schr\"{o}dinger equation shown in Fig.\ \ref{fig:2D}.}
\label{fig:2Dinteractions}
\end{figure}

This is additionally illustrated in Extended Data Fig.\ \ref{fig:2Dinteractions}, where the recovered zero-momentum fraction in the square lattice showed no change around $\tau \approx \SI{30}{\micro s}$ for small interactions (up to $a=20\,a_0$).

The recovered zero-momentum fractions at $\tau = \SI{250}{\micro s}$ shown in Fig.\ \ref{fig:interactions} in the main text are therefore sufficient to determine the slope of the whole curve, which we use as an indicator of localisation (as shown in Fig. \ref{fig:4D}).

The chemical potential was calculated using the following expression valid within the Thomas-Fermi approximation \cite{l._p._pitaevskii_s._stringari_bose-einstein_2016}:
\begin{equation}
\mu = \frac{\hbar \bar{\omega}}{2} \left ( \frac{15 N a}{\bar{a}_{\mathrm{ho}}} \right )^{2/5},
\end{equation}
where $a$ is the scattering length, $\bar{\omega}$ the geometric mean of the dipole trap frequencies, $N$ the atom number and $\bar{a}_{\mathrm{ho}} = \sqrt{\nicefrac{\hbar}{m \bar{\omega}}}$ the mean harmonic oscillator length.

The data in Fig.\ \ref{fig:interactions} is fitted phenomenologically with an error function:
\begin{equation} \label{eq:errfunc}
a\cdot\mathrm{Erf}[b(\mu-c)] + d, \quad \begin{tabular}{c|c|c|c|}
 & value & error\\
\hline
$a$ & 0.40 & 0.01 \\
\hline
$b$ & $5.72\times 10^{-3}$ & $2.9\times 10^{-4}$\\
\hline
$c$ & 290.0 & 5.7\\ 
\hline
$d$ & 0.58 & 0.01\\ 
\hline
\end{tabular}.
\end{equation}


\begin{thebibliography}{32}%
\makeatletter
\providecommand \@ifxundefined [1]{%
 \@ifx{#1\undefined}
}%
\providecommand \@ifnum [1]{%
 \ifnum #1\expandafter \@firstoftwo
 \else \expandafter \@secondoftwo
 \fi
}%
\providecommand \@ifx [1]{%
 \ifx #1\expandafter \@firstoftwo
 \else \expandafter \@secondoftwo
 \fi
}%
\providecommand \natexlab [1]{#1}%
\providecommand \enquote  [1]{``#1''}%
\providecommand \bibnamefont  [1]{#1}%
\providecommand \bibfnamefont [1]{#1}%
\providecommand \citenamefont [1]{#1}%
\providecommand \href@noop [0]{\@secondoftwo}%
\providecommand \href [0]{\begingroup \@sanitize@url \@href}%
\providecommand \@href[1]{\@@startlink{#1}\@@href}%
\providecommand \@@href[1]{\endgroup#1\@@endlink}%
\providecommand \@sanitize@url [0]{\catcode `\\12\catcode `\$12\catcode
  `\&12\catcode `\#12\catcode `\^12\catcode `\_12\catcode `\%12\relax}%
\providecommand \@@startlink[1]{}%
\providecommand \@@endlink[0]{}%
\providecommand \url  [0]{\begingroup\@sanitize@url \@url }%
\providecommand \@url [1]{\endgroup\@href {#1}{\urlprefix }}%
\providecommand \urlprefix  [0]{URL }%
\providecommand \Eprint [0]{\href }%
\providecommand \doibase [0]{http://dx.doi.org/}%
\providecommand \selectlanguage [0]{\@gobble}%
\providecommand \bibinfo  [0]{\@secondoftwo}%
\providecommand \bibfield  [0]{\@secondoftwo}%
\providecommand \translation [1]{[#1]}%
\providecommand \BibitemOpen [0]{}%
\providecommand \bibitemStop [0]{}%
\providecommand \bibitemNoStop [0]{.\EOS\space}%
\providecommand \EOS [0]{\spacefactor3000\relax}%
\providecommand \BibitemShut  [1]{\csname bibitem#1\endcsname}%
\let\auto@bib@innerbib\@empty
\bibitem [{\citenamefont {Senechal}(1995)}]{senechal_quasicrystals_1995}%
  \BibitemOpen
  \bibfield  {author} {\bibinfo {author} {\bibfnamefont {Marjorie}\
  \bibnamefont {Senechal}},\ }\href@noop {} {\emph {\bibinfo {title}
  {Quasicrystals and geometry}}}\ (\bibinfo  {publisher} {Cambridge University
  Press},\ \bibinfo {address} {Cambridge},\ \bibinfo {year} {1995})\BibitemShut
  {NoStop}%
\bibitem [{\citenamefont {Puig}(2006)}]{asch_cantor_2006}%
  \BibitemOpen
  \bibfield  {author} {\bibinfo {author} {\bibfnamefont {Joaquim}\ \bibnamefont
  {Puig}},\ }\bibfield  {title} {{\selectlanguage {english}\enquote {\bibinfo
  {title} {Cantor {Spectrum} for {Quasi}-{Periodic} {Schrödinger}
  {Operators}},}\ }}in\ \href {\doibase 10.1007/3-540-34273-7_9}
  {{\selectlanguage {english}\emph {\bibinfo {booktitle} {Mathematical
  {Physics} of {Quantum} {Mechanics}}}}},\ Vol.\ \bibinfo {volume} {690},\
  \bibinfo {editor} {edited by\ \bibinfo {editor} {\bibfnamefont {Joachim}\
  \bibnamefont {Asch}}\ and\ \bibinfo {editor} {\bibfnamefont {Alain}\
  \bibnamefont {Joye}}}\ (\bibinfo  {publisher} {Springer Berlin Heidelberg},\
  \bibinfo {year} {2006})\ pp.\ \bibinfo {pages} {79--91}\BibitemShut {NoStop}%
\bibitem [{\citenamefont {Ahn}\ \emph {et~al.}(2018)\citenamefont {Ahn},
  \citenamefont {Moon}, \citenamefont {Kim}, \citenamefont {Kim}, \citenamefont
  {Shin}, \citenamefont {Kim}, \citenamefont {Cha}, \citenamefont {Kahng},
  \citenamefont {Kim}, \citenamefont {Koshino}, \citenamefont {Son},
  \citenamefont {Yang},\ and\ \citenamefont {Ahn}}]{ahn_dirac_2018}%
  \BibitemOpen
  \bibfield  {author} {\bibinfo {author} {\bibfnamefont {Sung~Joon}\
  \bibnamefont {Ahn}}, \bibinfo {author} {\bibfnamefont {Pilkyung}\
  \bibnamefont {Moon}}, \bibinfo {author} {\bibfnamefont {Tae-Hoon}\
  \bibnamefont {Kim}}, \bibinfo {author} {\bibfnamefont {Hyun-Woo}\
  \bibnamefont {Kim}}, \bibinfo {author} {\bibfnamefont {Ha-Chul}\ \bibnamefont
  {Shin}}, \bibinfo {author} {\bibfnamefont {Eun~Hye}\ \bibnamefont {Kim}},
  \bibinfo {author} {\bibfnamefont {Hyun~Woo}\ \bibnamefont {Cha}}, \bibinfo
  {author} {\bibfnamefont {Se-Jong}\ \bibnamefont {Kahng}}, \bibinfo {author}
  {\bibfnamefont {Philip}\ \bibnamefont {Kim}}, \bibinfo {author}
  {\bibfnamefont {Mikito}\ \bibnamefont {Koshino}}, \bibinfo {author}
  {\bibfnamefont {Young-Woo}\ \bibnamefont {Son}}, \bibinfo {author}
  {\bibfnamefont {Cheol-Woong}\ \bibnamefont {Yang}}, \ and\ \bibinfo {author}
  {\bibfnamefont {Joung~Real}\ \bibnamefont {Ahn}},\ }\bibfield  {title}
  {{\selectlanguage {english}\enquote {\bibinfo {title} {Dirac electrons in a
  dodecagonal graphene quasicrystal},}\ }}\href {\doibase
  10.1126/science.aar8412} {\bibfield  {journal} {\bibinfo  {journal}
  {Science}\ }\textbf {\bibinfo {volume} {361}},\ \bibinfo {pages} {782--786}
  (\bibinfo {year} {2018})}\BibitemShut {NoStop}%
\bibitem [{\citenamefont {Vardeny}\ \emph {et~al.}(2013)\citenamefont
  {Vardeny}, \citenamefont {Nahata},\ and\ \citenamefont
  {Agrawal}}]{vardeny_optics_2013}%
  \BibitemOpen
  \bibfield  {author} {\bibinfo {author} {\bibfnamefont {Z.~Valy}\ \bibnamefont
  {Vardeny}}, \bibinfo {author} {\bibfnamefont {Ajay}\ \bibnamefont {Nahata}},
  \ and\ \bibinfo {author} {\bibfnamefont {Amit}\ \bibnamefont {Agrawal}},\
  }\bibfield  {title} {\enquote {\bibinfo {title} {Optics of photonic
  quasicrystals},}\ }\href {https://doi.org/10.1038/nphoton.2012.343}
  {\bibfield  {journal} {\bibinfo  {journal} {Nature Photonics}\ }\textbf
  {\bibinfo {volume} {7}},\ \bibinfo {pages} {177} (\bibinfo {year}
  {2013})}\BibitemShut {NoStop}%
\bibitem [{\citenamefont {Guidoni}\ \emph {et~al.}(1997)\citenamefont
  {Guidoni}, \citenamefont {Triché}, \citenamefont {Verkerk},\ and\
  \citenamefont {Grynberg}}]{guidoni_quasiperiodic_1997}%
  \BibitemOpen
  \bibfield  {author} {\bibinfo {author} {\bibfnamefont {L.}~\bibnamefont
  {Guidoni}}, \bibinfo {author} {\bibfnamefont {C.}~\bibnamefont {Triché}},
  \bibinfo {author} {\bibfnamefont {P.}~\bibnamefont {Verkerk}}, \ and\
  \bibinfo {author} {\bibfnamefont {G.}~\bibnamefont {Grynberg}},\ }\bibfield
  {title} {\enquote {\bibinfo {title} {Quasiperiodic {Optical} {Lattices}},}\
  }\href {\doibase 10.1103/PhysRevLett.79.3363} {\bibfield  {journal} {\bibinfo
   {journal} {Phys. Rev. Lett.}\ }\textbf {\bibinfo {volume} {79}},\ \bibinfo
  {pages} {3363--3366} (\bibinfo {year} {1997})}\BibitemShut {NoStop}%
\bibitem [{\citenamefont {Roati}\ \emph {et~al.}(2008)\citenamefont {Roati},
  \citenamefont {D’Errico}, \citenamefont {Fallani}, \citenamefont {Fattori},
  \citenamefont {Fort}, \citenamefont {Zaccanti}, \citenamefont {Modugno},
  \citenamefont {Modugno},\ and\ \citenamefont
  {Inguscio}}]{roati_anderson_2008}%
  \BibitemOpen
  \bibfield  {author} {\bibinfo {author} {\bibfnamefont {Giacomo}\ \bibnamefont
  {Roati}}, \bibinfo {author} {\bibfnamefont {Chiara}\ \bibnamefont
  {D’Errico}}, \bibinfo {author} {\bibfnamefont {Leonardo}\ \bibnamefont
  {Fallani}}, \bibinfo {author} {\bibfnamefont {Marco}\ \bibnamefont
  {Fattori}}, \bibinfo {author} {\bibfnamefont {Chiara}\ \bibnamefont {Fort}},
  \bibinfo {author} {\bibfnamefont {Matteo}\ \bibnamefont {Zaccanti}}, \bibinfo
  {author} {\bibfnamefont {Giovanni}\ \bibnamefont {Modugno}}, \bibinfo
  {author} {\bibfnamefont {Michele}\ \bibnamefont {Modugno}}, \ and\ \bibinfo
  {author} {\bibfnamefont {Massimo}\ \bibnamefont {Inguscio}},\ }\bibfield
  {title} {\enquote {\bibinfo {title} {Anderson localization of a
  non-interacting {Bose}–{Einstein} condensate},}\ }\href {\doibase
  10.1038/nature07071} {\bibfield  {journal} {\bibinfo  {journal} {Nature}\
  }\textbf {\bibinfo {volume} {453}},\ \bibinfo {pages} {895--898} (\bibinfo
  {year} {2008})}\BibitemShut {NoStop}%
\bibitem [{\citenamefont {Gadway}\ \emph {et~al.}(2011)\citenamefont {Gadway},
  \citenamefont {Pertot}, \citenamefont {Reeves}, \citenamefont {Vogt},\ and\
  \citenamefont {Schneble}}]{gadway_glassy_2011}%
  \BibitemOpen
  \bibfield  {author} {\bibinfo {author} {\bibfnamefont {Bryce}\ \bibnamefont
  {Gadway}}, \bibinfo {author} {\bibfnamefont {Daniel}\ \bibnamefont {Pertot}},
  \bibinfo {author} {\bibfnamefont {Jeremy}\ \bibnamefont {Reeves}}, \bibinfo
  {author} {\bibfnamefont {Matthias}\ \bibnamefont {Vogt}}, \ and\ \bibinfo
  {author} {\bibfnamefont {Dominik}\ \bibnamefont {Schneble}},\ }\bibfield
  {title} {{\selectlanguage {english}\enquote {\bibinfo {title} {Glassy
  behavior in a binary atomic mixture},}\ }}\href {\doibase
  10.1103/PhysRevLett.107.145306} {\bibfield  {journal} {\bibinfo  {journal}
  {Physical Review Letters}\ }\textbf {\bibinfo {volume} {107}},\ \bibinfo
  {pages} {145306} (\bibinfo {year} {2011})}\BibitemShut {NoStop}%
\bibitem [{\citenamefont {Schreiber}\ \emph {et~al.}(2015)\citenamefont
  {Schreiber}, \citenamefont {Hodgman}, \citenamefont {Bordia}, \citenamefont
  {Luschen}, \citenamefont {Fischer}, \citenamefont {Vosk}, \citenamefont
  {Altman}, \citenamefont {Schneider},\ and\ \citenamefont
  {Bloch}}]{schreiber_observation_2015}%
  \BibitemOpen
  \bibfield  {author} {\bibinfo {author} {\bibfnamefont {M.}~\bibnamefont
  {Schreiber}}, \bibinfo {author} {\bibfnamefont {S.~S.}\ \bibnamefont
  {Hodgman}}, \bibinfo {author} {\bibfnamefont {P.}~\bibnamefont {Bordia}},
  \bibinfo {author} {\bibfnamefont {H.~P.}\ \bibnamefont {Luschen}}, \bibinfo
  {author} {\bibfnamefont {M.~H.}\ \bibnamefont {Fischer}}, \bibinfo {author}
  {\bibfnamefont {R.}~\bibnamefont {Vosk}}, \bibinfo {author} {\bibfnamefont
  {E.}~\bibnamefont {Altman}}, \bibinfo {author} {\bibfnamefont
  {U.}~\bibnamefont {Schneider}}, \ and\ \bibinfo {author} {\bibfnamefont
  {I.}~\bibnamefont {Bloch}},\ }\bibfield  {title} {{\selectlanguage
  {english}\enquote {\bibinfo {title} {Observation of many-body localization of
  interacting fermions in a quasirandom optical lattice},}\ }}\href {\doibase
  10.1126/science.aaa7432} {\bibfield  {journal} {\bibinfo  {journal}
  {Science}\ }\textbf {\bibinfo {volume} {349}},\ \bibinfo {pages} {842--845}
  (\bibinfo {year} {2015})}\BibitemShut {NoStop}%
\bibitem [{\citenamefont {Rajagopal}\ \emph {et~al.}(2019)\citenamefont
  {Rajagopal}, \citenamefont {Shimasaki}, \citenamefont {Dotti}, \citenamefont
  {Račiūnas}, \citenamefont {Senaratne}, \citenamefont {Anisimovas},
  \citenamefont {Eckardt},\ and\ \citenamefont
  {Weld}}]{rajagopal_phasonic_2019}%
  \BibitemOpen
  \bibfield  {author} {\bibinfo {author} {\bibfnamefont {Shankari~V.}\
  \bibnamefont {Rajagopal}}, \bibinfo {author} {\bibfnamefont {Toshihiko}\
  \bibnamefont {Shimasaki}}, \bibinfo {author} {\bibfnamefont {Peter}\
  \bibnamefont {Dotti}}, \bibinfo {author} {\bibfnamefont {Mantas}\
  \bibnamefont {Račiūnas}}, \bibinfo {author} {\bibfnamefont {Ruwan}\
  \bibnamefont {Senaratne}}, \bibinfo {author} {\bibfnamefont {Egidijus}\
  \bibnamefont {Anisimovas}}, \bibinfo {author} {\bibfnamefont {André}\
  \bibnamefont {Eckardt}}, \ and\ \bibinfo {author} {\bibfnamefont {David~M.}\
  \bibnamefont {Weld}},\ }\bibfield  {title} {{\selectlanguage
  {english}\enquote {\bibinfo {title} {Phasonic {Spectroscopy} of a {Quantum}
  {Gas} in a {Quasicrystalline} {Lattice}},}\ }}\href {\doibase
  10.1103/PhysRevLett.123.223201} {\bibfield  {journal} {\bibinfo  {journal}
  {Physical Review Letters}\ }\textbf {\bibinfo {volume} {123}},\ \bibinfo
  {pages} {223201} (\bibinfo {year} {2019})}\BibitemShut {NoStop}%
\bibitem [{\citenamefont {Cubitt}\ \emph {et~al.}(2015)\citenamefont {Cubitt},
  \citenamefont {Perez-Garcia},\ and\ \citenamefont
  {Wolf}}]{cubitt_undecidability_2015}%
  \BibitemOpen
  \bibfield  {author} {\bibinfo {author} {\bibfnamefont {Toby~S.}\ \bibnamefont
  {Cubitt}}, \bibinfo {author} {\bibfnamefont {David}\ \bibnamefont
  {Perez-Garcia}}, \ and\ \bibinfo {author} {\bibfnamefont {Michael~M.}\
  \bibnamefont {Wolf}},\ }\bibfield  {title} {\enquote {\bibinfo {title}
  {Undecidability of the spectral gap},}\ }\href
  {https://doi.org/10.1038/nature16059} {\bibfield  {journal} {\bibinfo
  {journal} {Nature}\ }\textbf {\bibinfo {volume} {528}},\ \bibinfo {pages}
  {207} (\bibinfo {year} {2015})}\BibitemShut {NoStop}%
\bibitem [{\citenamefont {Fisher}\ \emph {et~al.}(1989)\citenamefont {Fisher},
  \citenamefont {Weichman}, \citenamefont {Grinstein},\ and\ \citenamefont
  {Fisher}}]{fisher_boson_1989}%
  \BibitemOpen
  \bibfield  {author} {\bibinfo {author} {\bibfnamefont {Matthew P.~A.}\
  \bibnamefont {Fisher}}, \bibinfo {author} {\bibfnamefont {Peter~B.}\
  \bibnamefont {Weichman}}, \bibinfo {author} {\bibfnamefont {G.}~\bibnamefont
  {Grinstein}}, \ and\ \bibinfo {author} {\bibfnamefont {Daniel~S.}\
  \bibnamefont {Fisher}},\ }\bibfield  {title} {\enquote {\bibinfo {title}
  {Boson localization and the superfluid-insulator transition},}\ }\href
  {\doibase 10.1103/PhysRevB.40.546} {\bibfield  {journal} {\bibinfo  {journal}
  {Phys. Rev. B}\ }\textbf {\bibinfo {volume} {40}},\ \bibinfo {pages}
  {546--570} (\bibinfo {year} {1989})}\BibitemShut {NoStop}%
\bibitem [{\citenamefont {Söyler}\ \emph {et~al.}(2011)\citenamefont
  {Söyler}, \citenamefont {Kiselev}, \citenamefont {Prokofev},\ and\
  \citenamefont {Svistunov}}]{soyler_phase_2011}%
  \BibitemOpen
  \bibfield  {author} {\bibinfo {author} {\bibfnamefont {S}~\bibnamefont
  {Söyler}}, \bibinfo {author} {\bibfnamefont {M.}~\bibnamefont {Kiselev}},
  \bibinfo {author} {\bibfnamefont {N.~V.}\ \bibnamefont {Prokofev}}, \ and\
  \bibinfo {author} {\bibfnamefont {B.~V.}\ \bibnamefont {Svistunov}},\
  }\bibfield  {title} {\enquote {\bibinfo {title} {Phase {Diagram} of the
  {Commensurate} {Two}-{Dimensional} {Disordered} {Bose}-{Hubbard} {Model}},}\
  }\href {\doibase 10.1103/PhysRevLett.107.185301} {\bibfield  {journal}
  {\bibinfo  {journal} {Phys. Rev. Lett.}\ }\textbf {\bibinfo {volume} {107}},\
  \bibinfo {pages} {185301} (\bibinfo {year} {2011})}\BibitemShut {NoStop}%
\bibitem [{\citenamefont {Bardarson}\ \emph {et~al.}(2017)\citenamefont
  {Bardarson}, \citenamefont {Pollmann}, \citenamefont {Schneider},\ and\
  \citenamefont {Sondhi}}]{Andp2017}%
  \BibitemOpen
  \bibfield  {author} {\bibinfo {author} {\bibfnamefont {Jens~H.}\ \bibnamefont
  {Bardarson}}, \bibinfo {author} {\bibfnamefont {Frank}\ \bibnamefont
  {Pollmann}}, \bibinfo {author} {\bibfnamefont {Ulrich}\ \bibnamefont
  {Schneider}}, \ and\ \bibinfo {author} {\bibfnamefont {Shivaji}\ \bibnamefont
  {Sondhi}},\ }\bibfield  {title} {\enquote {\bibinfo {title} {Special issue:
  Many‐body localization},}\ }\href
  {https://onlinelibrary.wiley.com/doi/abs/10.1002/andp.201700191} {\bibfield
  {journal} {\bibinfo  {journal} {Annalen der Physik}\ }\textbf {\bibinfo
  {volume} {529}} (\bibinfo {year} {2017})}\BibitemShut {NoStop}%
\bibitem [{\citenamefont {Choi}\ \emph {et~al.}(2016)\citenamefont {Choi},
  \citenamefont {Hild}, \citenamefont {Zeiher}, \citenamefont {Schau{\ss}},
  \citenamefont {Rubio-Abadal}, \citenamefont {Yefsah}, \citenamefont
  {Khemani}, \citenamefont {Huse}, \citenamefont {Bloch},\ and\ \citenamefont
  {Gross}}]{Choi2016}%
  \BibitemOpen
  \bibfield  {author} {\bibinfo {author} {\bibfnamefont {Jae-yoon}\
  \bibnamefont {Choi}}, \bibinfo {author} {\bibfnamefont {Sebastian}\
  \bibnamefont {Hild}}, \bibinfo {author} {\bibfnamefont {Johannes}\
  \bibnamefont {Zeiher}}, \bibinfo {author} {\bibfnamefont {Peter}\
  \bibnamefont {Schau{\ss}}}, \bibinfo {author} {\bibfnamefont {Antonio}\
  \bibnamefont {Rubio-Abadal}}, \bibinfo {author} {\bibfnamefont {Tarik}\
  \bibnamefont {Yefsah}}, \bibinfo {author} {\bibfnamefont {Vedika}\
  \bibnamefont {Khemani}}, \bibinfo {author} {\bibfnamefont {David~A.}\
  \bibnamefont {Huse}}, \bibinfo {author} {\bibfnamefont {Immanuel}\
  \bibnamefont {Bloch}}, \ and\ \bibinfo {author} {\bibfnamefont {Christian}\
  \bibnamefont {Gross}},\ }\bibfield  {title} {\enquote {\bibinfo {title}
  {Exploring the many-body localization transition in two dimensions},}\ }\href
  {\doibase 10.1126/science.aaf8834} {\bibfield  {journal} {\bibinfo  {journal}
  {Science}\ }\textbf {\bibinfo {volume} {352}},\ \bibinfo {pages} {1547--1552}
  (\bibinfo {year} {2016})}\BibitemShut {NoStop}%
\bibitem [{\citenamefont {Bordia}\ \emph
  {et~al.}(2017{\natexlab{a}})\citenamefont {Bordia}, \citenamefont
  {L\"uschen}, \citenamefont {Scherg}, \citenamefont {Gopalakrishnan},
  \citenamefont {Knap}, \citenamefont {Schneider},\ and\ \citenamefont
  {Bloch}}]{Bordia2D}%
  \BibitemOpen
  \bibfield  {author} {\bibinfo {author} {\bibfnamefont {Pranjal}\ \bibnamefont
  {Bordia}}, \bibinfo {author} {\bibfnamefont {Henrik}\ \bibnamefont
  {L\"uschen}}, \bibinfo {author} {\bibfnamefont {Sebastian}\ \bibnamefont
  {Scherg}}, \bibinfo {author} {\bibfnamefont {Sarang}\ \bibnamefont
  {Gopalakrishnan}}, \bibinfo {author} {\bibfnamefont {Michael}\ \bibnamefont
  {Knap}}, \bibinfo {author} {\bibfnamefont {Ulrich}\ \bibnamefont
  {Schneider}}, \ and\ \bibinfo {author} {\bibfnamefont {Immanuel}\
  \bibnamefont {Bloch}},\ }\bibfield  {title} {\enquote {\bibinfo {title}
  {Probing slow relaxation and many-body localization in two-dimensional
  quasiperiodic systems},}\ }\href {\doibase 10.1103/PhysRevX.7.041047}
  {\bibfield  {journal} {\bibinfo  {journal} {Phys. Rev. X}\ }\textbf {\bibinfo
  {volume} {7}},\ \bibinfo {pages} {041047} (\bibinfo {year}
  {2017}{\natexlab{a}})}\BibitemShut {NoStop}%
\bibitem [{\citenamefont {De~Roeck}\ and\ \citenamefont
  {Huveneers}(2017)}]{de_roeck_stability_2017}%
  \BibitemOpen
  \bibfield  {author} {\bibinfo {author} {\bibfnamefont {Wojciech}\
  \bibnamefont {De~Roeck}}\ and\ \bibinfo {author} {\bibfnamefont {François}\
  \bibnamefont {Huveneers}},\ }\bibfield  {title} {\enquote {\bibinfo {title}
  {Stability and instability towards delocalization in many-body localization
  systems},}\ }\href {\doibase 10.1103/PhysRevB.95.155129} {\bibfield
  {journal} {\bibinfo  {journal} {Phys. Rev. B}\ }\textbf {\bibinfo {volume}
  {95}},\ \bibinfo {pages} {155129} (\bibinfo {year} {2017})}\BibitemShut
  {NoStop}%
\bibitem [{\citenamefont {Potirniche}\ \emph {et~al.}(2019)\citenamefont
  {Potirniche}, \citenamefont {Banerjee},\ and\ \citenamefont
  {Altman}}]{Potirniche2019}%
  \BibitemOpen
  \bibfield  {author} {\bibinfo {author} {\bibfnamefont {Ionut~Dragos}\
  \bibnamefont {Potirniche}}, \bibinfo {author} {\bibfnamefont {Sumilan}\
  \bibnamefont {Banerjee}}, \ and\ \bibinfo {author} {\bibfnamefont {Ehud}\
  \bibnamefont {Altman}},\ }\bibfield  {title} {\enquote {\bibinfo {title}
  {{Exploration of the stability of many-body localization in
  d{\textgreater}1}},}\ }\href@noop {} {\bibfield  {journal} {\bibinfo
  {journal} {Physical Review B}\ }\textbf {\bibinfo {volume} {99}} (\bibinfo
  {year} {2019})}\BibitemShut {NoStop}%
\bibitem [{\citenamefont {Szab{\'{o}}}\ and\ \citenamefont
  {Schneider}(2020)}]{Szabo2020}%
  \BibitemOpen
  \bibfield  {author} {\bibinfo {author} {\bibfnamefont {Attila}\ \bibnamefont
  {Szab{\'{o}}}}\ and\ \bibinfo {author} {\bibfnamefont {Ulrich}\ \bibnamefont
  {Schneider}},\ }\bibfield  {title} {\enquote {\bibinfo {title} {{Mixed
  spectra and partially extended states in a two-dimensional quasiperiodic
  model}},}\ }\href {\doibase 10.1103/PhysRevB.101.014205} {\bibfield
  {journal} {\bibinfo  {journal} {Physical Review B}\ }\textbf {\bibinfo
  {volume} {101}},\ \bibinfo {pages} {014205} (\bibinfo {year}
  {2020})}\BibitemShut {NoStop}%
\bibitem [{\citenamefont {Shechtman}\ \emph {et~al.}(1984)\citenamefont
  {Shechtman}, \citenamefont {Blech}, \citenamefont {Gratias},\ and\
  \citenamefont {Cahn}}]{shechtman_metallic_1984}%
  \BibitemOpen
  \bibfield  {author} {\bibinfo {author} {\bibfnamefont {D.}~\bibnamefont
  {Shechtman}}, \bibinfo {author} {\bibfnamefont {I.}~\bibnamefont {Blech}},
  \bibinfo {author} {\bibfnamefont {D.}~\bibnamefont {Gratias}}, \ and\
  \bibinfo {author} {\bibfnamefont {J.~W.}\ \bibnamefont {Cahn}},\ }\bibfield
  {title} {\enquote {\bibinfo {title} {Metallic {Phase} with {Long}-{Range}
  {Orientational} {Order} and {No} {Translational} {Symmetry}},}\ }\href
  {\doibase 10.1103/PhysRevLett.53.1951} {\bibfield  {journal} {\bibinfo
  {journal} {Phys. Rev. Lett.}\ }\textbf {\bibinfo {volume} {53}},\ \bibinfo
  {pages} {1951--1953} (\bibinfo {year} {1984})}\BibitemShut {NoStop}%
\bibitem [{\citenamefont {Viebahn}\ \emph {et~al.}(2019)\citenamefont
  {Viebahn}, \citenamefont {Sbroscia}, \citenamefont {Carter}, \citenamefont
  {Yu},\ and\ \citenamefont {Schneider}}]{viebahn_matter-wave_2019}%
  \BibitemOpen
  \bibfield  {author} {\bibinfo {author} {\bibfnamefont {Konrad}\ \bibnamefont
  {Viebahn}}, \bibinfo {author} {\bibfnamefont {Matteo}\ \bibnamefont
  {Sbroscia}}, \bibinfo {author} {\bibfnamefont {Edward}\ \bibnamefont
  {Carter}}, \bibinfo {author} {\bibfnamefont {Jr-Chiun}\ \bibnamefont {Yu}}, \
  and\ \bibinfo {author} {\bibfnamefont {Ulrich}\ \bibnamefont {Schneider}},\
  }\bibfield  {title} {\enquote {\bibinfo {title} {Matter-{Wave} {Diffraction}
  from a {Quasicrystalline} {Optical} {Lattice}},}\ }\href {\doibase
  10.1103/PhysRevLett.122.110404} {\bibfield  {journal} {\bibinfo  {journal}
  {Phys. Rev. Lett.}\ }\textbf {\bibinfo {volume} {122}},\ \bibinfo {pages}
  {110404} (\bibinfo {year} {2019})}\BibitemShut {NoStop}%
\bibitem [{noa()}]{noauthor_notitle_nodate-1}%
  \BibitemOpen
  \href@noop {} {\ }\bibinfo {note} {See supplemental material}\BibitemShut
  {NoStop}%
\bibitem [{\citenamefont {Denschlag}\ \emph {et~al.}(2002)\citenamefont
  {Denschlag}, \citenamefont {Simsarian}, \citenamefont {Häffner},
  \citenamefont {McKenzie}, \citenamefont {Browaeys}, \citenamefont {Cho},
  \citenamefont {Helmerson}, \citenamefont {Rolston},\ and\ \citenamefont
  {Phillips}}]{denschlag_bose-einstein_2002}%
  \BibitemOpen
  \bibfield  {author} {\bibinfo {author} {\bibfnamefont {J.~Hecker}\
  \bibnamefont {Denschlag}}, \bibinfo {author} {\bibfnamefont {J.~E.}\
  \bibnamefont {Simsarian}}, \bibinfo {author} {\bibfnamefont {H.}~\bibnamefont
  {Häffner}}, \bibinfo {author} {\bibfnamefont {C.}~\bibnamefont {McKenzie}},
  \bibinfo {author} {\bibfnamefont {A.}~\bibnamefont {Browaeys}}, \bibinfo
  {author} {\bibfnamefont {D.}~\bibnamefont {Cho}}, \bibinfo {author}
  {\bibfnamefont {K.}~\bibnamefont {Helmerson}}, \bibinfo {author}
  {\bibfnamefont {S.~L.}\ \bibnamefont {Rolston}}, \ and\ \bibinfo {author}
  {\bibfnamefont {W.~D.}\ \bibnamefont {Phillips}},\ }\bibfield  {title}
  {\enquote {\bibinfo {title} {A {Bose}-{Einstein} condensate in an optical
  lattice},}\ }\href {\doibase 10.1088/0953-4075/35/14/307} {\bibfield
  {journal} {\bibinfo  {journal} {Journal of Physics B: Atomic, Molecular and
  Optical Physics}\ }\textbf {\bibinfo {volume} {35}},\ \bibinfo {pages}
  {3095--3110} (\bibinfo {year} {2002})}\BibitemShut {NoStop}%
\bibitem [{\citenamefont {Szab\'o}\ and\ \citenamefont
  {Schneider}(2018)}]{Szabo2018}%
  \BibitemOpen
  \bibfield  {author} {\bibinfo {author} {\bibfnamefont {Attila}\ \bibnamefont
  {Szab\'o}}\ and\ \bibinfo {author} {\bibfnamefont {Ulrich}\ \bibnamefont
  {Schneider}},\ }\bibfield  {title} {\enquote {\bibinfo {title} {Non-power-law
  universality in one-dimensional quasicrystals},}\ }\href {\doibase
  10.1103/PhysRevB.98.134201} {\bibfield  {journal} {\bibinfo  {journal} {Phys.
  Rev. B}\ }\textbf {\bibinfo {volume} {98}},\ \bibinfo {pages} {134201}
  (\bibinfo {year} {2018})}\BibitemShut {NoStop}%
\bibitem [{\citenamefont {Fletcher}\ \emph {et~al.}(2017)\citenamefont
  {Fletcher}, \citenamefont {Lopes}, \citenamefont {Man}, \citenamefont
  {Navon}, \citenamefont {Smith}, \citenamefont {Zwierlein},\ and\
  \citenamefont {Hadzibabic}}]{fletcher_two-_2017}%
  \BibitemOpen
  \bibfield  {author} {\bibinfo {author} {\bibfnamefont {Richard~J.}\
  \bibnamefont {Fletcher}}, \bibinfo {author} {\bibfnamefont {Raphael}\
  \bibnamefont {Lopes}}, \bibinfo {author} {\bibfnamefont {Jay}\ \bibnamefont
  {Man}}, \bibinfo {author} {\bibfnamefont {Nir}\ \bibnamefont {Navon}},
  \bibinfo {author} {\bibfnamefont {Robert~P.}\ \bibnamefont {Smith}}, \bibinfo
  {author} {\bibfnamefont {Martin~W.}\ \bibnamefont {Zwierlein}}, \ and\
  \bibinfo {author} {\bibfnamefont {Zoran}\ \bibnamefont {Hadzibabic}},\
  }\bibfield  {title} {{\selectlanguage {english}\enquote {\bibinfo {title}
  {Two- and three-body contacts in the unitary {Bose} gas},}\ }}\href {\doibase
  10.1126/science.aai8195} {\bibfield  {journal} {\bibinfo  {journal}
  {Science}\ }\textbf {\bibinfo {volume} {355}},\ \bibinfo {pages} {377--380}
  (\bibinfo {year} {2017})}\BibitemShut {NoStop}%
\bibitem [{\citenamefont {Gericke}\ \emph {et~al.}(2007)\citenamefont
  {Gericke}, \citenamefont {Gerbier}, \citenamefont {Widera}, \citenamefont
  {Fölling}, \citenamefont {Mandel},\ and\ \citenamefont
  {Bloch}}]{gericke_adiabatic_2007}%
  \BibitemOpen
  \bibfield  {author} {\bibinfo {author} {\bibfnamefont {T.}~\bibnamefont
  {Gericke}}, \bibinfo {author} {\bibfnamefont {F.}~\bibnamefont {Gerbier}},
  \bibinfo {author} {\bibfnamefont {A.}~\bibnamefont {Widera}}, \bibinfo
  {author} {\bibfnamefont {S.}~\bibnamefont {Fölling}}, \bibinfo {author}
  {\bibfnamefont {O.}~\bibnamefont {Mandel}}, \ and\ \bibinfo {author}
  {\bibfnamefont {I.}~\bibnamefont {Bloch}},\ }\bibfield  {title} {\enquote
  {\bibinfo {title} {Adiabatic loading of a {Bose}–{Einstein} condensate in a
  {3D} optical lattice},}\ }\href {\doibase 10.1080/09500340600777730}
  {\bibfield  {journal} {\bibinfo  {journal} {Journal of Modern Optics}\
  }\textbf {\bibinfo {volume} {54}},\ \bibinfo {pages} {735--743} (\bibinfo
  {year} {2007})}\BibitemShut {NoStop}%
\bibitem [{\citenamefont {Sornette}(1998)}]{sornette_discrete-scale_1998}%
  \BibitemOpen
  \bibfield  {author} {\bibinfo {author} {\bibfnamefont {Didier}\ \bibnamefont
  {Sornette}},\ }\bibfield  {title} {\enquote {\bibinfo {title} {Discrete-scale
  invariance and complex dimensions},}\ }\href {\doibase
  https://doi.org/10.1016/S0370-1573(97)00076-8} {\bibfield  {journal}
  {\bibinfo  {journal} {Physics Reports}\ }\textbf {\bibinfo {volume} {297}},\
  \bibinfo {pages} {239 -- 270} (\bibinfo {year} {1998})}\BibitemShut {NoStop}%
\bibitem [{\citenamefont {Eckardt}(2017)}]{Eckardt2017}%
  \BibitemOpen
  \bibfield  {author} {\bibinfo {author} {\bibfnamefont {Andr{\'{e}}}\
  \bibnamefont {Eckardt}},\ }\bibfield  {title} {\enquote {\bibinfo {title}
  {{Colloquium: Atomic quantum gases in periodically driven optical
  lattices}},}\ }\href@noop {} {\bibfield  {journal} {\bibinfo  {journal}
  {Reviews of Modern Physics}\ }\textbf {\bibinfo {volume} {89}} (\bibinfo
  {year} {2017})}\BibitemShut {NoStop}%
\bibitem [{\citenamefont {Bordia}\ \emph
  {et~al.}(2017{\natexlab{b}})\citenamefont {Bordia}, \citenamefont {Lüschen},
  \citenamefont {Schneider}, \citenamefont {Knap},\ and\ \citenamefont
  {Bloch}}]{bordia_periodically_2017}%
  \BibitemOpen
  \bibfield  {author} {\bibinfo {author} {\bibfnamefont {Pranjal}\ \bibnamefont
  {Bordia}}, \bibinfo {author} {\bibfnamefont {Henrik}\ \bibnamefont
  {Lüschen}}, \bibinfo {author} {\bibfnamefont {Ulrich}\ \bibnamefont
  {Schneider}}, \bibinfo {author} {\bibfnamefont {Michael}\ \bibnamefont
  {Knap}}, \ and\ \bibinfo {author} {\bibfnamefont {Immanuel}\ \bibnamefont
  {Bloch}},\ }\bibfield  {title} {\enquote {\bibinfo {title} {Periodically
  driving a many-body localized quantum system},}\ }\href
  {https://doi.org/10.1038/nphys4020} {\bibfield  {journal} {\bibinfo
  {journal} {Nature Physics}\ }\textbf {\bibinfo {volume} {13}},\ \bibinfo
  {pages} {460} (\bibinfo {year} {2017}{\natexlab{b}})}\BibitemShut {NoStop}%
\bibitem [{\citenamefont {Ozawa}\ and\ \citenamefont
  {Price}(2019)}]{Ozawa2019}%
  \BibitemOpen
  \bibfield  {author} {\bibinfo {author} {\bibfnamefont {Tomoki}\ \bibnamefont
  {Ozawa}}\ and\ \bibinfo {author} {\bibfnamefont {Hannah~M.}\ \bibnamefont
  {Price}},\ }\bibfield  {title} {\enquote {\bibinfo {title} {{Topological
  quantum matter in synthetic dimensions}},}\ }\href@noop {} {\bibfield
  {journal} {\bibinfo  {journal} {Nature Reviews Physics}\ }\textbf {\bibinfo
  {volume} {1}},\ \bibinfo {pages} {349--357} (\bibinfo {year}
  {2019})}\BibitemShut {NoStop}%
\bibitem [{\citenamefont {Gaunt}(2015)}]{gaunt_degenerate_2015}%
  \BibitemOpen
  \bibfield  {author} {\bibinfo {author} {\bibfnamefont {A.L.}\ \bibnamefont
  {Gaunt}},\ }\href {https://books.google.co.uk/books?id=lZoyvwEACAAJ} {\emph
  {\bibinfo {title} {Degenerate {Bose} {Gases}: {Tuning} {Interactions} \&
  {Geometry}}}}\ (\bibinfo  {publisher} {University of Cambridge},\ \bibinfo
  {year} {2015})\ \bibinfo {note} {phD thesis}\BibitemShut {NoStop}%
\bibitem [{\citenamefont {Lieb}\ \emph {et~al.}(2002)\citenamefont {Lieb},
  \citenamefont {Seiringer},\ and\ \citenamefont
  {Yngvason}}]{lieb_superfluidity_2002}%
  \BibitemOpen
  \bibfield  {author} {\bibinfo {author} {\bibfnamefont {Elliott~H.}\
  \bibnamefont {Lieb}}, \bibinfo {author} {\bibfnamefont {Robert}\ \bibnamefont
  {Seiringer}}, \ and\ \bibinfo {author} {\bibfnamefont {Jakob}\ \bibnamefont
  {Yngvason}},\ }\bibfield  {title} {{\selectlanguage {english}\enquote
  {\bibinfo {title} {Superfluidity in dilute trapped {Bose} gases},}\ }}\href
  {https://link.aps.org/doi/10.1103/PhysRevB.66.134529} {\bibfield  {journal}
  {\bibinfo  {journal} {Physical Review B}\ }\textbf {\bibinfo {volume} {66}}
  (\bibinfo {year} {2002})}\BibitemShut {NoStop}%
\bibitem [{\citenamefont {{L. P. Pitaevskiĭ, S.
  Stringari}}(2016)}]{l._p._pitaevskii_s._stringari_bose-einstein_2016}%
  \BibitemOpen
  \bibfield  {author} {\bibinfo {author} {\bibnamefont {{L. P. Pitaevskiĭ, S.
  Stringari}}},\ }\href@noop {} {\emph {\bibinfo {title} {Bose-{Einstein}
  condensation and superfluidity}}}\ (\bibinfo  {publisher} {Oxford University
  Press},\ \bibinfo {address} {Oxford},\ \bibinfo {year} {2016})\BibitemShut
  {NoStop}%
\end{thebibliography}
\end{document}